\documentclass{article}

\usepackage{authblk}
\usepackage{amsmath}
\usepackage{amsfonts}
\usepackage{graphicx,subcaption}
\usepackage{url}
\title{Labyrinth Chaos: Revisiting the Elegant, Chaotic and Hyperchaotic Walks}

\author[1,2]{Vasileios Basios}
\author[3]{Chris G. Antonopoulos}
\author[4]{Anouchah Latifi}

\affil[1]{Service de Physique des Syst\`emes Complexes et M\'ecanique Statistique, Universit\'e Libre de Bruxelles, Belgium}

\affil[2]{Interdisciplinary Center for Nonlinear Phenomena and Complex Systems (CeNoLi), Belgium\thanks{E-mail: vbasios@ulb.ac.be}}

\affil[3]{Department of Mathematical Sciences, University of Essex, UK\thanks{E-mail: canton@essex.ac.uk}}

\affil[4]{Department of Physics, Faculty of Sciences, Qom University of Technology, Iran\thanks{E-mail: latifi@qut.ac.ir}}

\date{\today}

\begin{document}

\maketitle
\begin{abstract}
\noindent Labyrinth chaos was discovered by Otto R\"ossler and Ren\'e Thomas in their endeavour to identify the necessary mathematical conditions for the appearance of chaotic and hyperchaotic motion in continuous flows. Here, we celebrate their discovery by considering a single labyrinth walks system and an array of coupled labyrinth chaos systems that exhibit complex, chaotic behaviour, reminiscent of chimera-like states, a peculiar synchronisation phenomenon. We discuss the properties of the single labyrinth walks system and review the ability of coupled labyrinth chaos systems to exhibit chimera-like states due to the unique properties of their space-filling, chaotic trajectories, what amounts to elegant, hyperchaotic walks. Finally, we discuss further implications in relation to the labyrinth walks system by showing that even though it is volume-preserving, it is not force-conservative.
\end{abstract}

\vskip 1cm

\vskip 1cm

\noindent\textbf{In the course of their pioneering work on the properties of feedback circuits and their relation to chaos and hyperchaos, Otto R\"ossler and Ren\'e Thomas proposed a minimal model of a dynamical system they termed Labyrinth Chaos. It turned out that even though it is simple, it is full of surprising properties that no-one could thought of. Simple and elegant as it is, it still holds great promise in elucidating aspects of chaotic dynamics that are not evident in other systems. Our paper revisits their work and highlights the incredible riches of this system in its disconcerting simplicity and importance in the context of dynamical systems and in other fields.
}

\section{Introduction}\label{Sec:Intro}

Undoubtedly, two of the most celebrated, low-dimensional, chaotic systems in the theory of dynamical systems are the R\"ossler and Lorenz systems with their iconic R\"ossler and Lorenz attractors \cite{Topologychaos,Gilmore:2008}. R\"ossler's work to elucidate the fundamentals of the topology of chaos, led him to investigate expanding and contracting circuits for his famous folding mechanism. His construction remains one of the simplest and most elegant chaotic dynamical systems to date, the R\"ossler system \cite{SprottBook}. 

R\"ossler's imperative to focus on the role of feedback, bore fruits quickly: For the first time, the R\"ossler system provided a clear demonstration of one of the two scenaria for the Shilnikov-route to chaos, the period doubling cascade \cite{Gaspard:1983c}. In his work, in bringing forward the concept of feedback circuits in dynamical systems theory, he found a great ally and coworker, Thomas. Thomas was already a renown biologist and deeply familiar with the idea of feedback circuits and their Boolean-logic aspects, present in phenomena associated with gene-expression, homeostasis, multi-stationarity and memory. He joined forces with his friend R\"ossler and turned him to an ``amateur mathematician'', as he would put it! R\"ossler, wrote in \cite{Letellier:2006} about his attractor and the role of feedback circuits: ``Later, Ren\'e Thomas saw much deeper into the topology of this feedback circuit''. However, the truth is that together, they managed to see so deep that they discovered many things which are fundamental in chaotic dynamics. Among them, was the importance of the coexistence of positive and negative circuits as the necessary mathematical conditions for the appearance of chaos and hyperchaos, in the sense of the presence of more than one positive Lyapunov exponents. In this context, the term ``circuits'' was used by R\"ossler and Thomas to refer to those terms in the Jacobian matrix of the dynamical system, whose row and column indices are in circular permutation. Therefore, circuits can be positive or negative according to the sign of the product of their terms as they appear in the determinant of the Jacobian matrix \cite{Thomas1999, ThomRoss2004} of the system.

In their venture \cite{Thomas1999, ThomRoss2004, ThomRoss2006}, R\"ossler and Thomas proposed a class of continuous-flow dynamical systems, called Thomas-R\"ossler (TR) systems \cite{BasAnt2019}. Their motive was to investigate what kind of nonlinearity is necessary to support chaos and hyperchaos in the simplest possible way. This led them to focus on the necessary mathematical conditions for the appearance of chaos, in terms of positive and negative circuits.

Their pioneering work on TR systems provided the ground for a plethora of published works of biological relevance (e.g. the special issue in memory of Thomas \cite{KaufmanJTBspecial}, and contributions therein) and of a more general scope in complex systems \cite{NicNic, KaufmanJTBspecial}, especially related to the emergence of complex behaviour via simple-circuit feedback structures in complex-systems science at large.

The main results of their work can be summarised in the following three statements: (i) a positive circuit is necessary for the system to have stable states, (ii) a negative circuit is necessary for the system to exhibit, robust, sustained oscillations, and (iii) a necessary condition for chaos is the presence of both a positive and a negative circuit in the system (for a more complete analysis, one may have a look at \cite{KaufmanJTBspecial}, where the authors discuss the subject in more details). However, concerning hyperchaos, we can add a fourth statement: (iv) a necessary condition for hyperchaos is the presence of more than one positive and negative, circuits.

What R\"ossler and Thomas called labyrinth chaos and ``Arabesques'' \cite{ThomRoss2004, Arabesque2013, BasAnt2019} drew wide attention as examples of elegant chaos, also discussed in \cite{SprottBook}. Here, the term ``elegant'' is understood in the sense of Poincar\'e's demand for beauty and simplicity in expressing mathematical ideas, something that R\"ossler and Thomas' work surely adheres to.

Interestingly, the class of TR systems exhibit typical properties, such as complex periodicity, multi-stationa\-rity, symmetries, etc. Notably, for appropriate parameter ranges, there typically appear coexisting, symmetric, chaotic attractors. However, one of their most spectacular findings is that of the special case of a totally new type of chaos in the absence of attractors: a chaotic, volume-preserving dynamical system with no attractors, the so-called labyrinth walks system. The existence of chaos in the absence of chaotic attractors is due to the existence of only unstable fixed points, located on an infinitely large, periodic lattice in a volume-preserving state-space. Hence, the trajectories wander through the lattice of unstable fixed points, exhibiting one or more positive Lyapunov exponents (i.e. hyperchaotic behaviour), depending on the dimensionality of the system. They termed these wandering, space filling trajectories, labyrinth walks \cite{Thomas1999, ThomRoss2004, ThomRoss2006}, in the sense they exhibit super-diffusive, fractional Brownian properties, which can also be described by complex, symbolic dynamics \cite{SprottChlouverakis}.

What has also been attributed to R\"ossler since the early dates of the development of chaos theory, is the term ``hyperchaos'' \cite{hyperchaos}, drawing attention to the topological idiosyncrasies of chaos, what refers to dynamics with more than one exponentially expanding directions, i.e. to more than one positive Lyapunov exponents. Ever since, its special significance has been studied in a plethora of works, ranging from fractional-order dynamical systems \cite{LI200455} to coupled quantum systems \cite{andreev2019chaos}. His contribution was and still remains seminal as certain fundamental questions, especially about the topology of chaos, are still open today \cite{chaos2020, Letellier:2007, Gilmore:2008}. R\"ossler and co-workers have also demonstrated that labyrinth chaos and labyrinth walks are hyperchaotic systems in more than three dimensions, showing the truth of condition (iv) as the necessary condition for the existence of hyperchaos \cite{ThomRoss2004}.

Amazingly, these systems still continue to surprise us with their elegance, beauty and combination of properties: As it has been reported recently in \cite{BasAnt2019}, arrays of coupled labyrinth chaos systems exhibit stereotypical, chimera-like states, reminiscent of chimera states observed in coupled Kuramoto oscillators \cite{Kuramotoetal2002,Abramsetal2004,Panaggioetal2015,BasAnt2019} and in other systems \cite{Hizanidisetal2016,Parasteshetal2018,Schmidtetal2017,erratumarticle,Sigalasetal2018,Kaneko2015}. These results emphasise further its significance as an example of an elegant dynamical system with a repertoire of rich properties, among which, hyperchaoticity is prominent.

The paper is organised as follows: In Sec. \ref{Sec:Circuits}, we review the concept of circuits introduced by R\"ossler and Thomas and their role in the appearance of chaos and hyperchaos in dynamical systems. In Sec. \ref{Sec:Hyperchaos}, we explore the dynamics in the neighbourhood of the fixed points of labyrinth walks and discuss how there can exist chaos and hyperchaos in the absence of attractors. Subsequently, in Sec. \ref{Sec:Chimeras}, we turn our attention to a more complex setting; namely, to that of an array of coupled TR systems and the role of labyrinth walks in the emergence of chimera-like states. In Sec. \ref{Sec:NonHam}, we discuss further implications in relation to the labyrinth walks system by showing 
that even though it is volume-preserving, it is not force-conservative. Finally, we conclude our work in Sec. \ref{Sec:Conclusion}, where we discuss possible future directions of work.

\section{Feedback Circuits and Chaos}\label{Sec:Circuits}
 
The concepts of circuits and feedback stem from electromechanical systems and biology. However, they were redefined by R\"ossler and Thomas to describe certain properties of dynamical systems and their equations of motion. We own them that ``complex dynamics can be thought of in terms of feedback circuits'' \cite{ThomRoss2004}. According to them, a feedback circuit (in short, a circuit) is the influence of a variable to its own evolution. This can happen directly or indirectly by the coupling of the variable with other variables in the system. Importantly, the coupling can be linear or nolinear and their realisation was pivotal in connecting graph theory with multistationarity and chaos \cite{Soule2003,Soule2007}.

In particular, what proved fruitful was that the feedback circuits are directly identifiable through the Jacobian matrix $J$ of a dynamical system, \[\frac{dX}{dt}=F(X),\] defined by
\begin{equation*}
J\equiv J(X)=\left[ \begin {array}{cccc} 
J_{11} & J_{12} & \cdots & J_{1n}\\
J_{21} & J_{22} & \cdots & J_{2n}\\
\vdots & \vdots & \ddots & \vdots\\
J_{n1} & J_{n2} & \cdots & J_{nn} 
\end{array}\right],
\end{equation*}
where $t$ is the time, $X=(X_1,\dots,X_n) \in\mathbb{R}^n$ the solution-vector and $F \colon\mathbb{R}^n\to\mathbb{R}^n$ a differentiable function with $F(X)=(F_1(X),\dots,F_n(X))$. The entries in the Jacobian matrix are the partial derivatives \[J_{ij}\equiv J_{ij}(X)=\frac{\partial F_i}{\partial X_j},\] where $i,j=1,\ldots,n$. 

Consequently, if $J_{ij}=0$, variable $X_j$ does not influence variable $X_i$. However, if $J_{ij}$ is different than zero, then variable $X_j$ influences variable $X_i$. This is in the sense that if $J_{ij}=\frac{\partial F_i}{\partial X_j}$ is positive, then there is a positive feedback from variable $X_j$ to variable $X_i$, whereas if $J_{ij}=\frac{\partial F_i}{\partial X_j}$ is negative, it results in a negative feedback from variable $X_j$ to variable $X_i$. Of course, the values of the entries in $J$ might depend on the solution $X(t)$.

\begin{figure}[h]
\centering{
\includegraphics[scale=0.65]{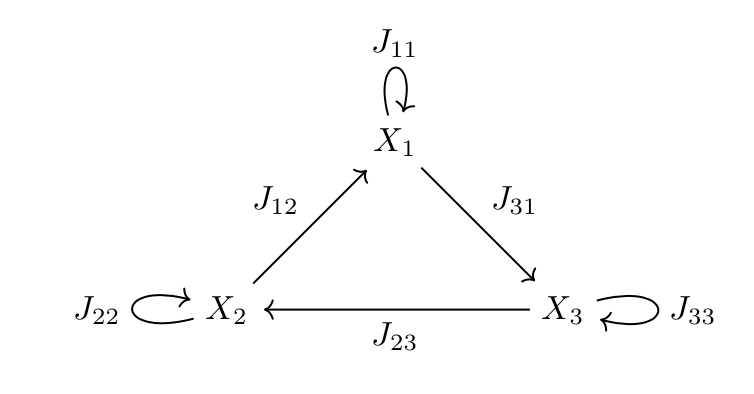}
}
%
%
%
%
%
\caption{Graph of incidences of the circuits in the Jacobian matrix $J$ of the system of Eq. \eqref{eq:LabN}, where $n=3$ and $b\neq0$. The curved arrows indicate self-loops (one-element circuits), where entries in $J$ influence themselves as they are its diagonal elements.}\label{f:diagram1}
\end{figure}

\begin{figure}[h]
\centering{
\includegraphics[scale=0.65]{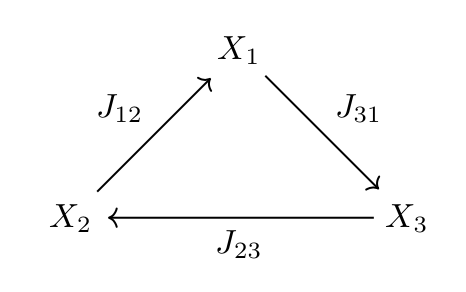}
}
%
%
%
\caption{Graph of incidences of the circuits in the Jacobian matrix $J$ of the system of Eq. \eqref{eq:3D-ThomRoss}, where $n=3$. Note the absence of self-loops as the diagonal elements in $J$ are equal to 0, thus there are no one-element circuits, in contrast to the graph of incidences of system \eqref{eq:LabN} in Fig. \ref{f:diagram1}.}\label{f:diagram2}
\end{figure}

Since a non-zero entry in $J$ implies that variable $X_j$ influences variable $X_i$, one can define a vertex from $X_j$ to $X_i$ in a graph of incidences (see Figs. \ref{f:diagram1} and \ref{f:diagram2} for two examples). Thus, circuits are the non-zero entries in $J$, with their indices forming cyclic permutations with each other.

For example, for the labyrinth chaos system of Eq. \eqref{eq:LabN} where $n=3$, the graph of incidences is shown in Fig. \ref{f:diagram1} for $b\neq0$, which shows its three element circuits, the closed path $X_1 \rightarrow X_3 \rightarrow X_2 \rightarrow X_1$ and the three, one-element, circuits which are self-loops, due to the diagonal elements in $J$.

In contrast, in Fig. \ref{f:diagram2}, the graph of incidences of the labyrinth walks system \eqref{eq:3D-ThomRoss}, where $n=3$ and $b=0$, has only one, three-element, circuit, which is the same as the three-element circuit in system \eqref{eq:LabN}, shown in Fig. \ref{f:diagram1}. However, there is no one-element circuit, since the diagonal elements in $J$ are zero, which implies there are no self-loops.

The circuits that can be identified in the Jacobian matrix are directly implicated in its characteristic equation, eigenvalues, eigenstates and determinant. Importantly, they are also implicated in the stability of fixed points and chaoticity, quantified by the spectrum of Lyapunov exponents.

Furthermore, circuits can be positive or negative depending on the sign of the product of their terms, i.e. on the signs of the entries in $J$. In this sense, the parity of the number of negative interactions of a circuit determines its sign: if it is even, the circuit is positive and if it is odd, it is negative \cite{ThomasNardone2009}. If its sign depends on the values of the solution $X$, i.e. on the position in the state-space, then the circuit is called an ambiguous circuit.

The locus of points that the sign of a non-ambiguous circuit changes from positive to negative is called a frontier. Depending of the character of change of the eigenvalues of $J$, the frontiers have been categorised in \cite{ThomasNardone2009}, where computer codes are also offered for their calculation.

The principal frontier \cite{ThomasNardone2009}, F1, is the locus of points where the sign of the determinant of the Jacobian matrix, $\det J$, changes and partitions the phase space according to the sign of the product of its eigenvalues. This is so as $ \prod \lambda_{k}=\det J$, where $\lambda_k$ are the eigenvalues of $J$.

Consequently, positive and negative circuits play contrasting roles in the dynamics of a system. The necessary condition for multistationarity is the presence of a positive circuit, whereas the presence of a negative circuit is a necessary condition for the existence of a stable attractor. Coexisting positive and negative circuits are necessary conditions for oscillations, chaos and hyperchaos (see conditions (i)-(iv)). Based on their work on TR systems and labyrinth walks, R\"ossler, Thomas and co-workers, proposed and demonstrated a necessary condition for the existence of hyperchaos: namely that hyperchaos of order $m$ (i.e. with $m$ positive Lyapunov exponents) can be generated by a single feedback circuit in $n=2m+1$ dimensions \cite{ThomRoss2004}.


The idea of feedback circuits is a common theme in engineering, biology, network science, nonlinear dynamics and complexity. The formulation of feedback circuits for the analysis of dynamical systems, as discussed by R\"ossler, Thomas, Rosen, Shilnikov, Nicolis and other pioneers in nonlinear dynamics and chaos theory, has become an integral part of the vocabulary, not only in relation to biological systems \cite{KaufmanJTBspecial} per se, but also, in complexity theory at large \cite{NicNic,JSNic,Kaufman2003complex}.

We shall now turn to the role of feedback circuits in the dynamics of the surprisingly elegant 3D labyrinth walks system.

\section{Chaos and Hyperchaos in the Absence of Attractors}\label{Sec:Hyperchaos}

We focus now on the system \cite{ThomRoss2004}
\begin{equation}\label{eq:LabN}
 \frac{dX_i}{dt}=\sin(X_{i+1}) - b X_i
\end{equation}
where $i=1,\dots,n$, $t$ is the time and $b$ a real constant. In this context, $X_{n+1}=X_1$, signifying the cyclic permutation in variables $X_i$. The system has $n$ one-element circuits, which makes it ideal for checking graphically, analytically and computationally the ideas developed by R\"ossler and Thomas. The particular case $b=0$ corresponds to the labyrinth walks system that we study next.

In particular, we study the case $n=3$ and $b=0$, what amounts to the 3D labyrinth walks system
\begin{equation}\label{eq:3D-ThomRoss}
\begin{aligned}
\frac{dX_1}{dt}&=\sin(X_2)=F_1,\\
\frac{dX_2}{dt}&=\sin(X_3)=F_2,\\
\frac{dX_3}{dt}&=\sin(X_1)=F_3,
\end{aligned}
\end{equation}
where $F=(F_1,F_2,F_3)$. System \eqref{eq:3D-ThomRoss} can be written in the compact form
\begin{equation*}
\dot{X}=P_M f(X),
\end{equation*}
where $X=(X_1,X_2,X_3)$, $\dot{X}=(\dot{X}_1,\dot{X}_2,\dot{X}_3)=\Bigl(\frac{dX_1}{dt},\frac{dX_2}{dt},\frac{dX_3}{dt}\Bigr)$,
\begin{equation*}
f(X)=\left[ \begin {array}{c} \sin\left(X_1\right)
\\\noalign{\medskip} \sin\left(X_2\right)
\\\noalign{\medskip}\sin \left( X_{3} \right)\end {array}
\right],
\end{equation*}
and
\begin{equation*}
P_M=\left[ \begin {array}{ccc} 0 & 1 &0
\\\noalign{\medskip}0& 0 &1 
\\\noalign{\medskip}1 &0& 0 \end {array}
\right]
\end{equation*}
is the permutation matrix.

The Jacobian matrix of system \eqref{eq:3D-ThomRoss} is given by
\begin{equation}\label{eq:3D-Jacobian}
\begin{aligned}
J&=\left[ \begin {array}{ccc} 0 &\cos \left( X_{2} \right) &0
\\\noalign{\medskip}0& 0 &\cos \left( X_{3} \right) 
\\\noalign{\medskip}\cos \left( X_{1} \right) &0& 0 \end {array}
\right]\\&=P_M\left[ \begin {array}{ccc} \dot{F_3} &0 &0
\\\noalign{\medskip}0& \dot{F_1} &0 
\\\noalign{\medskip}0&0& \dot{F_2} \end {array}
\right].
\end{aligned}
\end{equation}
Thus, the signs of circuit-changes occur in an alternating, nonlinear fashion depending on the cosines in $\dot{F}=(\dot{F_1},\dot{F_2},\dot{F_3})$, due to the cyclic permutations encoded in $P_M$. The determinant of $J$ is given by
\begin{equation*}
\det J(X)= \cos(X_1) \cos(X_2) \cos(X_3).
\end{equation*}

In the general case of $n$ variables, one can easily see that $J$ has the form of the permutation matrix $P_M$ in $n$ dimensions, where its entries are given by the generalisation of Eq. \eqref{eq:3D-Jacobian} in $n$ dimensions. It follows that in the $n$D case, its determinant is given by the product of the cosines of all variables, i.e. by
\begin{equation*}
\det J(X)=\prod_{i=1}^n\cos(X_i),
\end{equation*}
since the determinant of $P_M$ is 1 and the determinant of the corresponding diagonal matrix (see Eq. \eqref{eq:3D-Jacobian} for $n=3$) is $\prod_{i=1}^n \dot{F}_i=\prod_{i=1}^n \cos(X_i)$.
This shows that the principal frontier F1 of system \eqref{eq:3D-ThomRoss}, where the circuits change sign \cite{ThomasNardone2009}, has zero dimension. This is because F1 is an $nD$-lattice of points defined by the solution to the equation $\prod_{i=1}^n \cos(X_i)=0$. This is a set of distinct points and not a surface or hyper-surface, what is common in other systems \cite{ThomasNardone2009,Arabesque2013} studied by R\"ossler, Thomas and co-workers.
%
\begin{figure}[ht!]
\subfloat[\label{LabFixedPoints01}]{
\includegraphics[width=0.3\textwidth,height=0.3\textwidth,
]{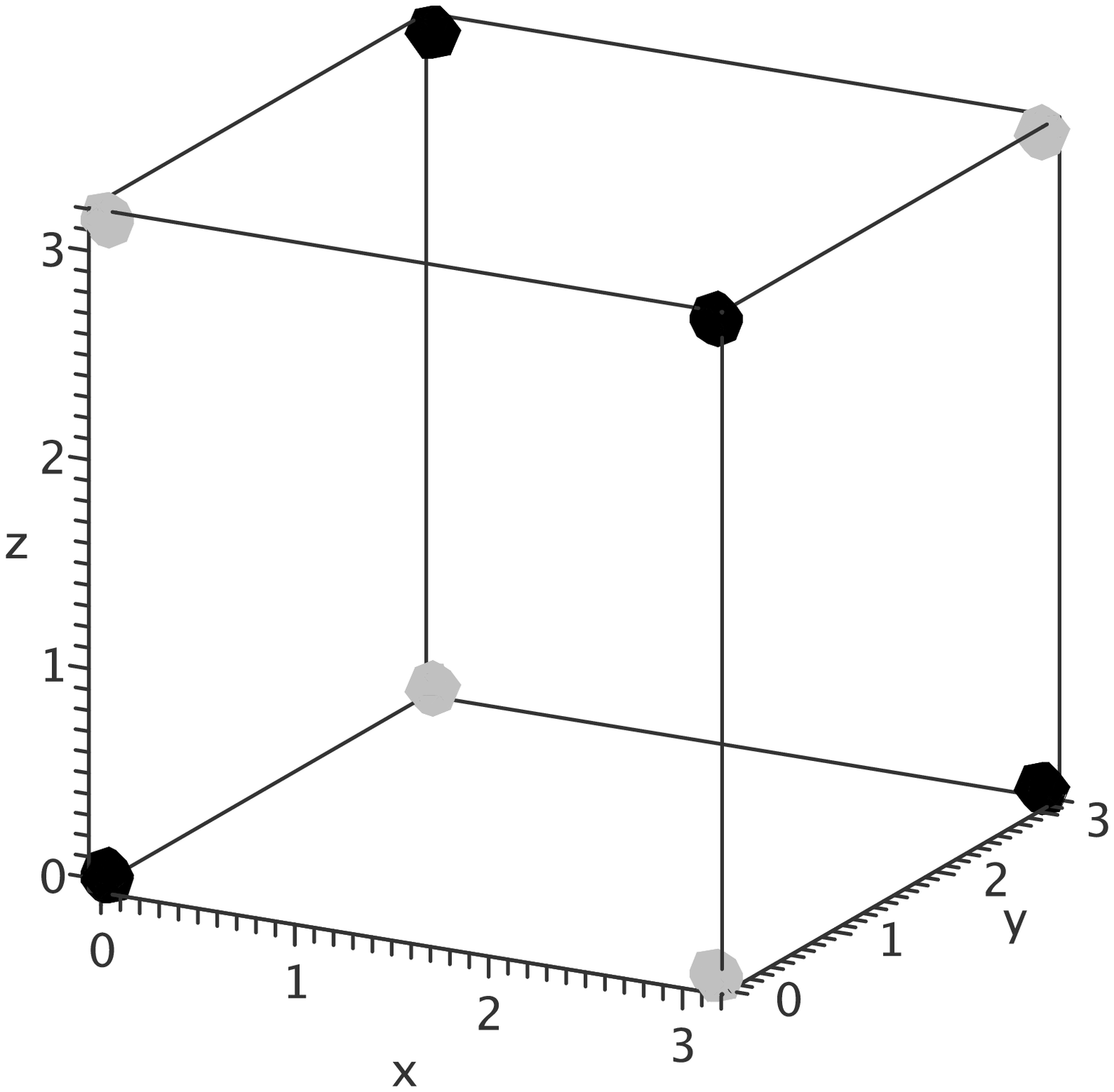}
}\hfill
\subfloat[\label{3DLattice}]{
\includegraphics[width=0.49\textwidth,height=0.39\textwidth,
]{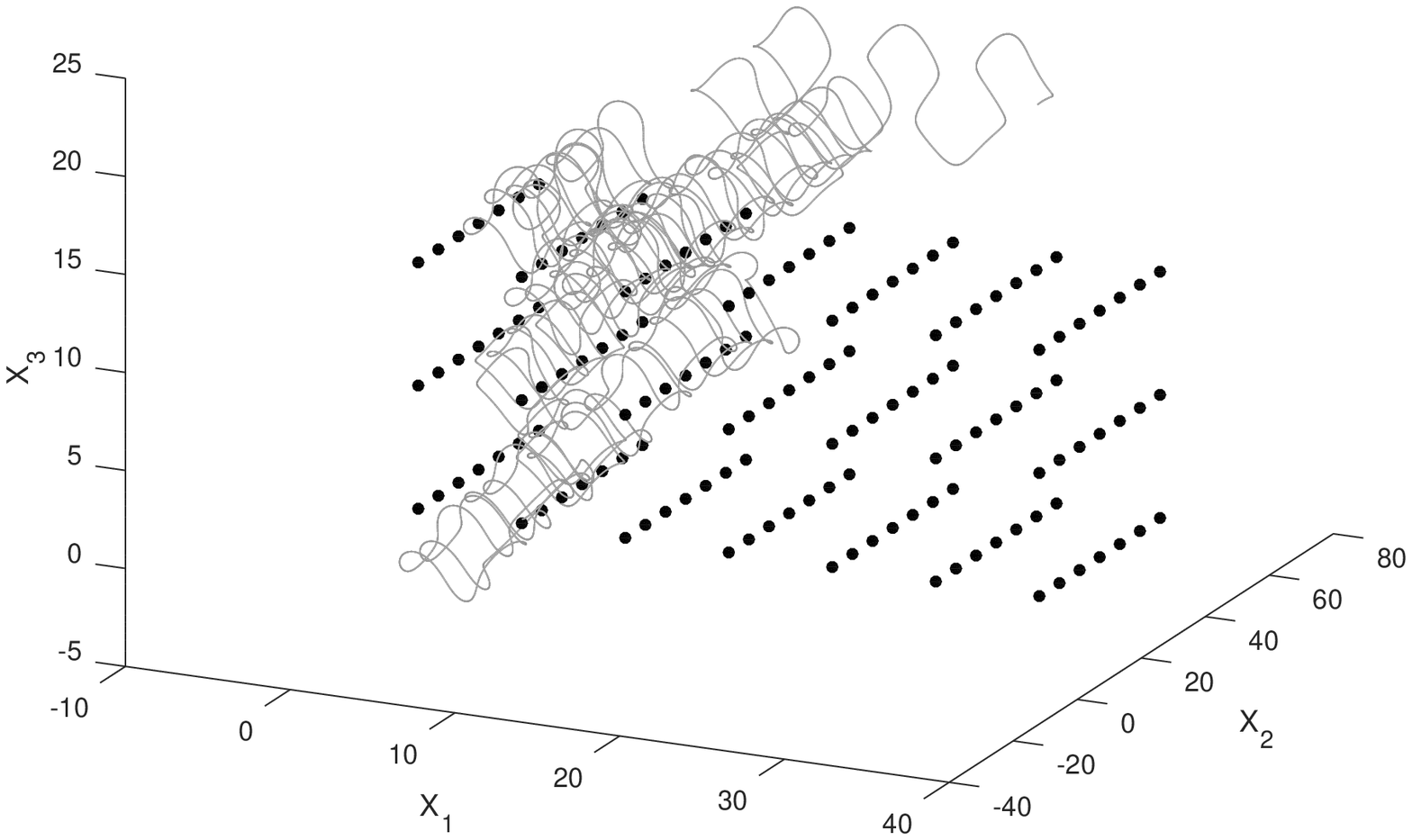}
}\hfill
\caption{Unit-cell of fixed points in panel (a) and part of the 3D fixed-points lattice with a labyrinth walk trajectory in panel (b). (a) The black and grey sets of unstable fixed points $X_i=k\pi,\;i=1,2,3$ of the labyrinth walks system \eqref{eq:3D-ThomRoss} in the unit cell for $k=0$ and 1. We note that the unit-length of the cell is $\pi$ in all 3 directions and that the state-space of system \eqref{eq:3D-ThomRoss} in panel (b) is composed of collated versions of the cell in (a) in all 3 directions. (b) Part of the lattice of fixed points in the 3D state-space of system \eqref{eq:3D-ThomRoss} (black points) and a labyrinth walks trajectory (grey curve).}\label{fig:lattice}
\end{figure}

The stability analysis of system \eqref{eq:3D-ThomRoss} shows there is a simple, yet elegant structure endowed with an intricate symmetry. The fixed points are the solutions to the equation $\frac{dX}{dt}=0$, therefore solutions to the system of equations
\begin{equation*}
\begin{aligned}
\sin(X_1)&=0,\\
\sin(X_2)&=0,\\
\sin(X_3)&=0.
\end{aligned}
\end{equation*}
This leads to a countably-infinite set of fixed points, arranged in the 3D lattice where $\sin(X_i) =0,\;i=1,2,3$, i.e. the fixed points are given by $X_i=k\pi,\;k\in\mathbb{N}$ (see Fig. \ref{fig:lattice}(b) for a subset of them). We note that these properties define an infinitely-large lattice of fixed points in the 3D state-space of the system composed of collated, identical cubic cells of unit-length $\pi$ (see Fig. \ref{fig:lattice}(a) for an example of such a cubic cell and (b) for a part of the infinitely-large 3D lattice of fixed points, what amounts to a number of collated, cubic cells of unit-length $\pi$). These properties can be carried out straightforwardly for any $n$. For example, for $k=0$ and 1, the set of fixed points of system \eqref{eq:3D-ThomRoss} is given by
$\{(0,0,0)$, 
$(0,0,\pi)$, 
$(0,\pi,0)$, 
$(\pi,0,0)$,
$(\pi,\pi,0)$,
$(\pi,0,\pi)$,
$(0,\pi,\pi)$,
$(\pi,\pi,\pi)\}$
and is shown in Fig. \ref{fig:lattice}(a). These are the only fixed points of the system and they all turn out to be unstable, what amounts to the absence of attractors. In particular, the eigenvalues of $J$ for each of the eight fixed points evaluate either to the cubic roots of unity, $\omega_{1,2,3}=\Big\{1, \frac{-1 \pm i\sqrt{3}}{2}\Big\}$, or to their opposites, $\omega_{1,2,3}=\Big\{-1, \frac{1 \mp i\sqrt{3}}{2}\Big\}$. Thus, the fixed points belong to two sets: The first set (black points in Fig. \ref{fig:lattice}(a)) consists of fixed points with one positive real eigenvalue and two complex, with negative real parts, eigenvalues, what signifies an outgoing unstable, spiral motion. 
The second set (grey points in Fig. \ref{fig:lattice}(a)) consists of fixed points with one negative real eigenvalue and two complex, with positive real parts, eigenvalues, what signifies an incoming, unstable, spiral motion. All fixed points, in both sets, have complex eigenvectors attributed to their complex eigenvalues, which adds a very interesting aspect for further investigations in terms of spinor properties \cite{ComplexEigen}. All fixed points are thus unstable and give rise to a structure that can sustain complex unstable periodic orbits, wandering, space-filling trajectories (the so-called, labyrinth walks) and thus, to chaotic behaviour.


\begin{figure}[ht!]
\subfloat[]{
\includegraphics[width=0.4\textwidth,height=0.49\textwidth,
]{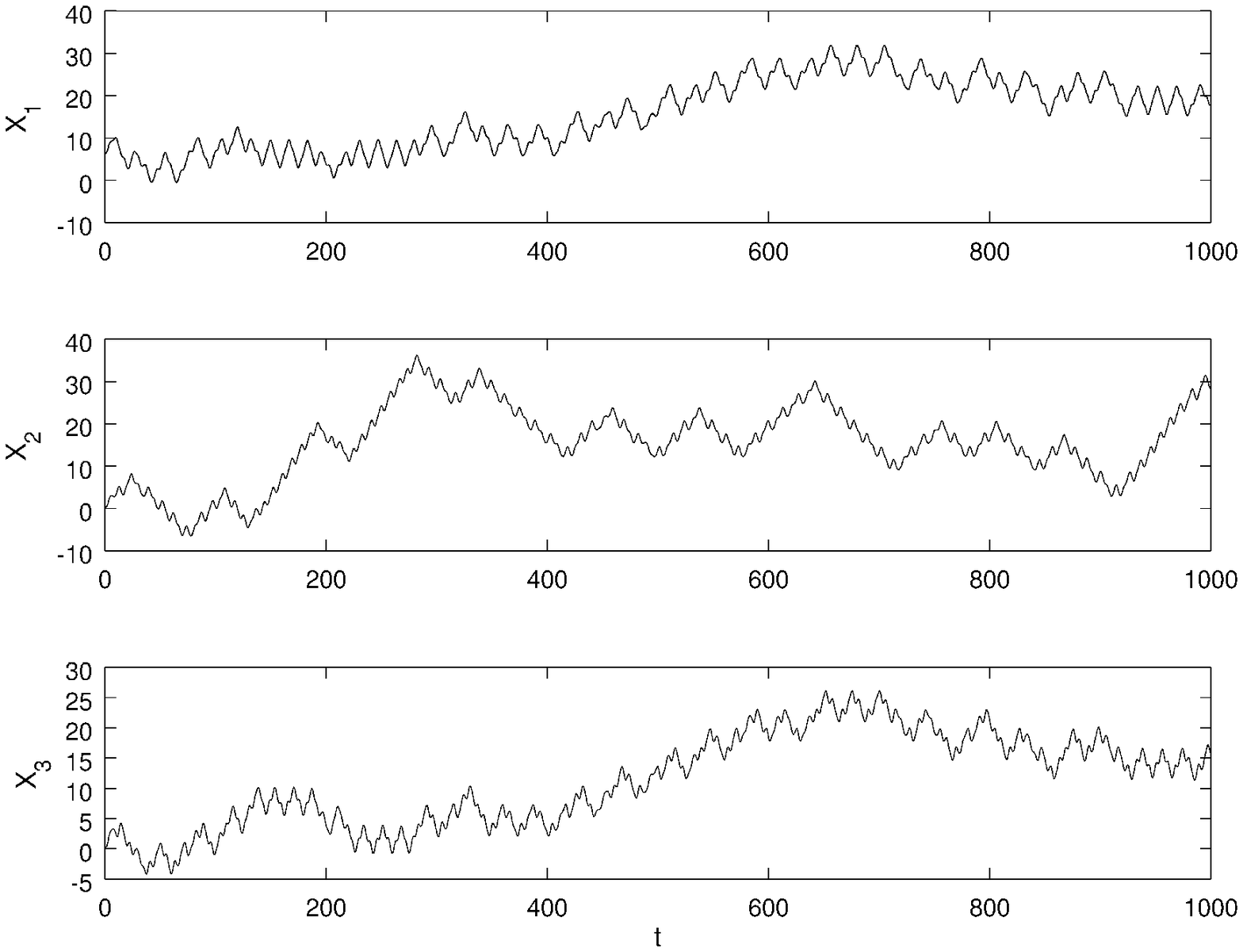}
}
\hspace{3.1cm}
\subfloat[\label{phasespace}]{
\includegraphics[width=0.4\textwidth,height=0.49\textwidth,
]{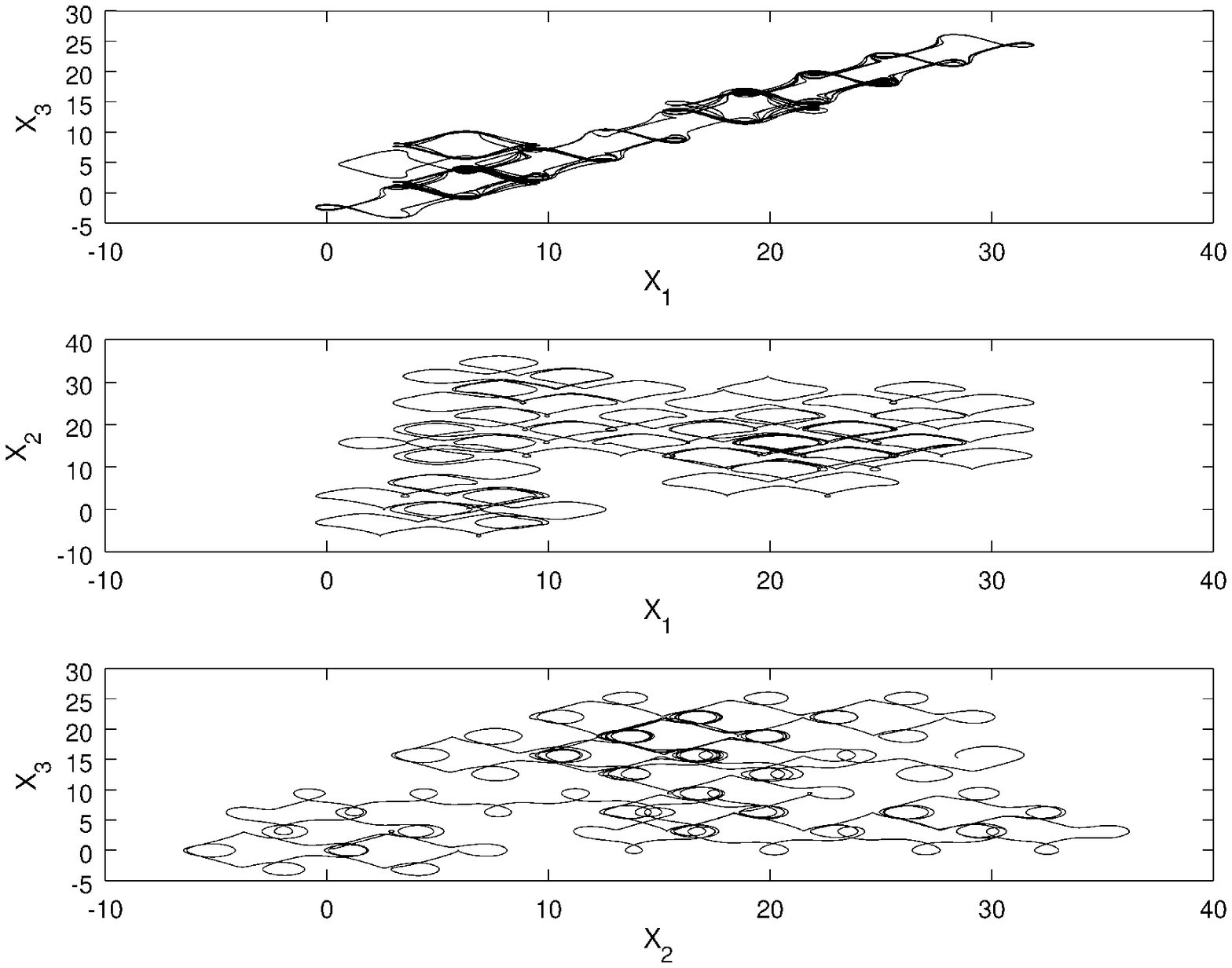}
}\hfill
\caption{Time evolution of $X_1$, $X_2$ and $X_3$ of the grey curve in Fig. \ref{fig:lattice}(b) and its projections in the $(X_3,X_1)$, $(X_2,X_1)$ and $(X_3,X_2)$ planes. (a) Evolution of $X_1$, $X_2$ and $X_3$ of the grey curve shown in Fig. \ref{fig:lattice}(b) over time, what constitutes an example of a labyrinth walks trajectory. (b) Projections of the same trajectory in the $(X_3,X_1)$, $(X_2,X_1)$ and $(X_3,X_2)$ planes.}\label{fig:TS-PhSp.eps}
\end{figure}

In Fig. \ref{fig:TS-PhSp.eps}(a), we present the time-evolution of $X_1$, $X_2$ and $X_3$ of system \eqref{eq:3D-ThomRoss} for the labyrinth walks, grey trajectory shown in Fig. \ref{fig:lattice}(b) and in Fig. \ref{fig:TS-PhSp.eps}(b), its projection in the planes $(X_3,X_1)$, $(X_2,X_1)$ and $(X_3,X_2)$, where some very interesting patterns can be seen. This trajectory is an example of a wandering, space-filling trajectory in the 3D state-space of system \eqref{eq:3D-ThomRoss}, part of which is shown in panel (b) in Fig. \ref{fig:lattice}, what amounts to the fractional, super-diffusive character \cite{SprottChlouverakis,SprottChlouverakis2007B} of labyrinth walks.

The results for the 3D case can be readily generalised in $n$ dimensions, where the unit cells will be $n$D hypercubes, building up an $n$D lattice of unstable fixed points, located at their vertices. In this case, the eigenvalues of the $n\times n$ Jacobian matrix $J$ will be the $n^{th}$ roots of unity, resulting in a structure of higher-order symmetries that can sustain higher-order hyperchaos without the presence of attractors. In \cite{ThomRoss2004}, it has been reported that systems with even and odd dimensionality have been examined and it has been predicted by R\"ossler and Thomas that they exhibit hyperchaos of order $m$, i.e. $m$ positive Lyapunov exponents, for $n=2m+1$. One-element circuits have been verified for $n=5,7,25,99$.

It is the existence of infinitely-many, unstable fixed points arranged on an infinite lattice, where there are no attractors that makes the labyrinth walks system so interesting. Even though it is deterministic, these properties lead to fractional-like chaotic motion reported in \cite{SprottChlouverakis,SprottChlouverakis2007B}, reminiscent of Levy-flights in stochastic systems! The coexistence of channels of free, ballistic flights along with unstable periodic orbits brings in mind trajectories and motion in infinite-horizon Lorenz-gas type billiards \cite{BilliardsLevy}. This combination of properties makes labyrinth walks not only an elegant system to study but also, a good example to elucidate the concept of volume preservation, a topic we will touch upon in Sec. \ref{Sec:NonHam}.

Before that, we turn our attention in the next section to coupled TR systems \eqref{eq:LabN} and revisit their surprising role in sustaining chimera-like states, a peculiar synchronisation phenomenon.

\section{Chimera-like States in arrays of Coupled Labyrinth Walks and TR systems}\label{Sec:Chimeras}
The fractional, super-diffusive character \cite{SprottChlouverakis,SprottChlouverakis2007B} of labyrinth walks provides the mechanism for the emergence of chimera-like states in arrays of coupled 3D TR systems \cite{BasAnt2019} (see Eqs. \eqref{eq:LabN}). In this section, we revisit the recent results in \cite{BasAnt2019}, in light of the ability of labyrinth walks and its lattice of unstable fixed points to facilitate complex, chimera-like states, even with very strong coupling.

To this end, the authors in \cite{BasAnt2019} considered the following system of coupled 3D labyrinth walks and TR systems arranged, in a simple ring topology with non-local coupling
\begin{equation}\label{eq:3DNPThomRoss}
\begin{aligned}
\frac{d X_1^{k}}{dt} &= -b_k X_1^{k} +\sin\big(X_2^{k}\big)+\frac{d}{2P}\sum_{j=k-P}^{k+P}\big(X_1^{k} - X_1^j\big),\\
\frac{d X_2^{k}}{dt} &= -b_k X_2^{k} + \sin\big(X_3^{k}\big),\\
\frac{d X_3^{k}}{dt} &= -b_k X_3^{k} + \sin\big(X_1^{k}\big),
\end{aligned}
\end{equation}
where $k=1,\dots,n$, $2P\leq n$ is the number of coupled neighbours, $d\geq 0$ is the strength of the linear coupling and $b_k\geq0$ for all $k$. The term $\frac{1}{2P}$ is a normalisation constant that distributes equally the linear coupling among the $n$ coupled systems.

\begin{figure}[!ht]
\subfloat[]{
\includegraphics[width=0.4\textwidth,height=0.45\textwidth,trim=0 0 10 6, clip]{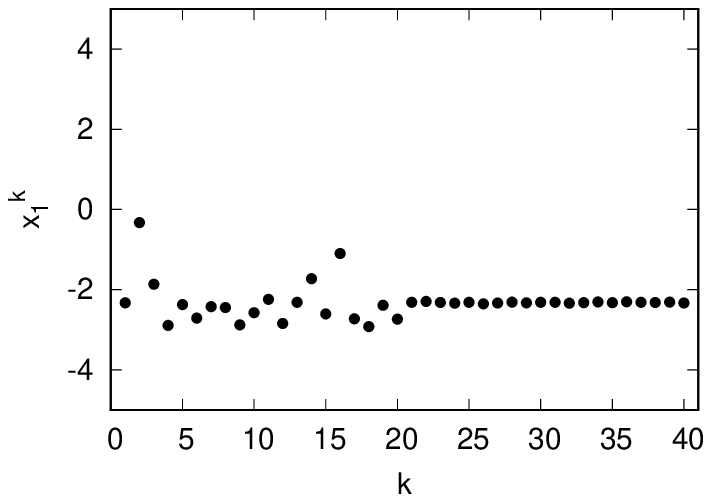}}
\hspace{2.75cm}
\subfloat[]{
\includegraphics[width=0.4\textwidth,height=0.45\textwidth,trim=0 0 10 6, clip]{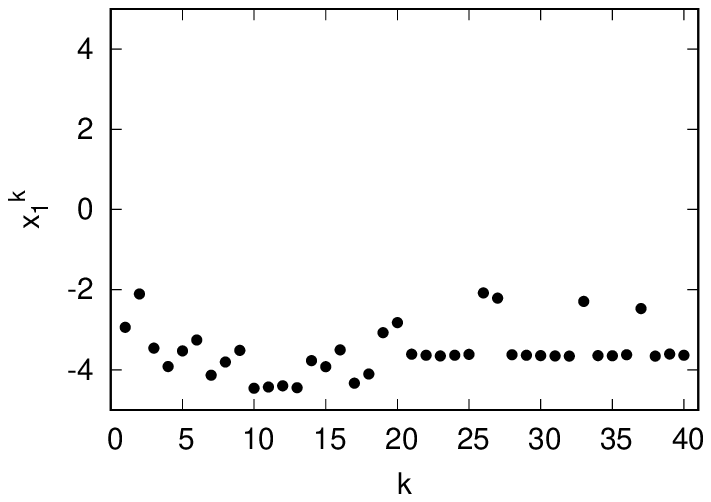}}
\caption{Snapshots of temporal phenomena of coherent and incoherent patterns, reminiscent of chimera-like states in system \eqref{eq:3DNPThomRoss} (see text for the details) as depicted in their spatio-temporal evolution in panel (b) 
in Fig. \ref{fig:spatiotemporal_plots_coupled_TR_systems}. (a) Snapshots at time $t\approx13401$ and (b) at a later time $t\approx13404$. Note that here, $d=0.6$, $n=40$, $b_k=0$ for $k=1,\ldots,20$ (labyrinth walks) and $b_k=0.17$ for $k=21,\ldots,40$ (labyrinth chaos). Note also that $X_1^k$ values seen at almost the same height (values at the vertical axes), imply they are locked (see text for more details). The group of locked $X_1^k$ values is the coherent group and the group of non-locked, $X_1^k$ values, the incoherent group.}\label{fig:chimera_like-states_coupled_TR_systems}
\end{figure}


\begin{figure}[!ht]
\subfloat[]{\vspace{3.6cm}
\includegraphics[width=0.48\textwidth,height=0.26\textwidth,trim=0 0 0 0, clip,angle=0]
{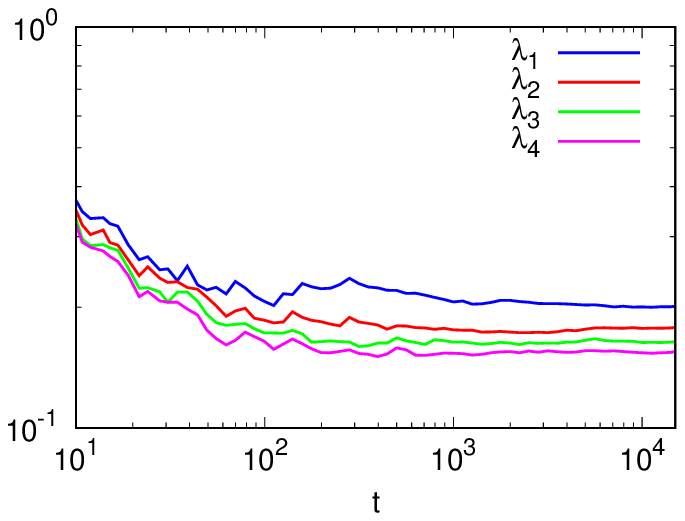}
}
\hspace{-0.28cm}
\subfloat[]{
\includegraphics[width=0.69\textwidth,height=0.48\textwidth,trim=31 60 30 158, clip,angle=-90]
{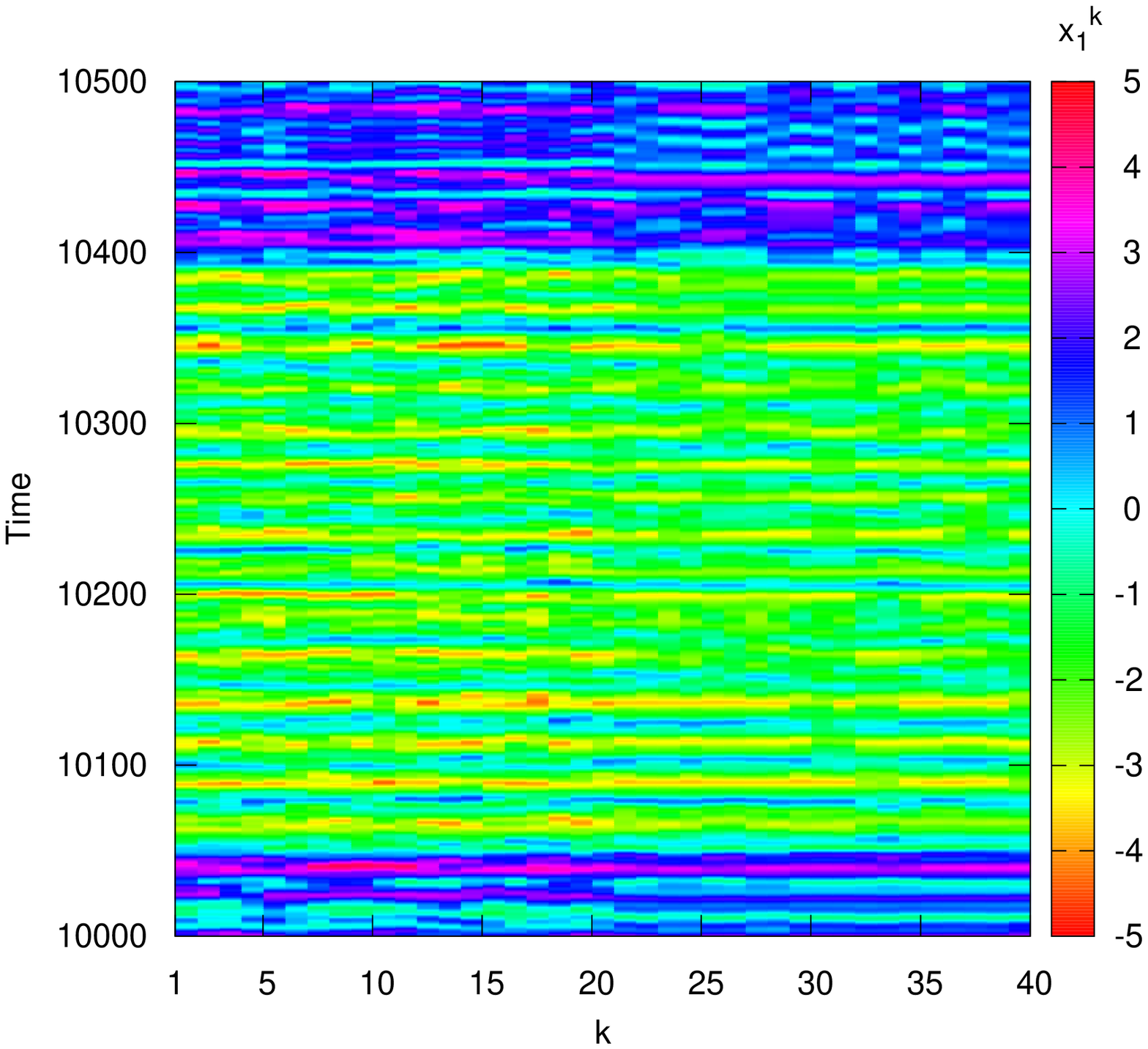}
}\hfill
\caption{Hyperchaotic and, spatio-temporal behaviour and emergence of alternating patterns of chimera-like states in system \eqref{eq:3DNPThomRoss}. (a) Evolution of the first four largest Lyapunov exponents $\lambda_i$, $i=1,\ldots,4$, seen to converge to positive, non-zero values (hyperchaoticity) and (b) chimera-like behaviour in $n=40$ TR and labyrinth walks systems \eqref{eq:3DNPThomRoss} (see text for the details). Note that here $d=0.6$, $b_k=0$ for $k=1,\ldots,20$ (labyrinth walks), $b_k=0.17$ for $k=21,\ldots,40$ (labyrinth chaos) and that the whole system is hyperchaotic with more than one positive Lyapunov exponents, as shown in panel (a).}\label{fig:spatiotemporal_plots_coupled_TR_systems}
\end{figure}

The system of Eqs. \eqref{eq:3DNPThomRoss} was introduced and studied in \cite{BasAnt2019}, where $n=40$ TR systems were considered, with the first half ($k=1,\ldots,20$) performing labyrinth walks ($b_k=0$) and the other half set in the labyrinth chaos regime with $b_k=0.18$ for $k=21,\ldots,40$. Each system was coupled with its $P=20$ nearest-neighbour systems in both sides (10 + 10), with the coupling strength set at $d=0.6$. For these parameters and setting, system \eqref{eq:3DNPThomRoss} is hyperchaotic, with more than 1 positive Lyapunov exponents.

For such a strong coupling ($d=0.6$) and without the presence of labyrinth walks in the second part of the system, one would expect the prevalence of chaotic synchronisation, what would result in an almost, fully synchronised system over time; yet, due to the presence of labyrinth walks, the emergence of partial, spatio-temporal synchronisation, reminiscent of chimera-like states \cite{Abramsetal2004,Panaggioetal2015,Hizanidisetal2016,Schmidtetal2017,Parasteshetal2018,Sigalasetal2018,Kaneko2015} was observed \cite{BasAnt2019}. In this context, partial is meant in the sense that the absolute differences of the $x_1^k$, $k=1,\ldots,n$ solutions were less than a threshold of the order of $10^{-4}$ to $10^{-3}$, i.e. they were locked in time, but not continuously over time. Here, we show in Figs. \ref{fig:chimera_like-states_coupled_TR_systems} and \ref{fig:spatiotemporal_plots_coupled_TR_systems}, the results of a similar study where the only difference with the study in \cite{BasAnt2019} is that $b_k=0.17$ for $k=21,\ldots,40$, which corresponds again to labyrinth chaos. In particular, Fig. \ref{fig:chimera_like-states_coupled_TR_systems}(a) shows an example of locking at two different times as depicted in the two panels for $t\approx13401$ and $t\approx13404$. Even though, these differences are locked (but not continuously) over time, the system remains hyperchaotic as a whole, 
as shown in Fig. \ref{fig:spatiotemporal_plots_coupled_TR_systems}(a), where the first four largest Lyapunov exponents $\lambda_i$, $i=1,\ldots,4$, can be seen to converge to positive, non-zero, values.

Apparently, it is the presence of chaotic walks that gives rise to the emergence of chimera-like states, what offers a novel phenomenon of chaotic synchronisation that is worth analysing further, both in terms of labyrinth walks and in terms of the appearance of chimera-like states in arrays of hyperchaotic dynamics. Moreover, these results are promising as they also offer the opportunity to study the emergence of chimera-like states in scale-free, small-world and Erd\H{o}s-R\'enyi networks of coupled TR systems.

\section{Labyrinth Walks: A Volume Preserving, non Force-conservative System}\label{Sec:NonHam}

We now turn to two very important properties of the labyrinth walks system \eqref{eq:3D-ThomRoss}.

The first one is that it is volume-preserving \cite{ODELiouville,AbrahamRalf}, in the sense it is preserving volumes in its 3D state-space. Indeed, if we write system \eqref{eq:3D-ThomRoss} in the form
\begin{equation*}
\dot{X}=F(X),
\end{equation*}
where, we recall, $X=(X_1,X_2,X_3)$ and $F=(F_1,F_2,F_3)=(\sin(X_2), \sin(X_3), \sin(X_1))$, the incompressibility condition (divergence) gives
\begin{equation}\label{eq:diveregence_condition}
\nabla\dot{X}=\nabla F=\frac{\partial F_1}{\partial X_1}+\frac{\partial F_2}{\partial X_2}+\frac{\partial F_3}{\partial X_3}=0.
\end{equation}
This means the flow $F$ of system \eqref{eq:3D-ThomRoss} is divergence-free, i.e. it is incompressible. This, in turn, implies that the solution to Liouville's equation \cite{Nolte2015}
\begin{equation*}
\frac{dV}{dt}=\int_V\big(\nabla\dot{X}\big)dV
\end{equation*}
is
\begin{equation*}
V(t)=e^{(\nabla\dot{X})t}V(0)=V(0),
\end{equation*}
where $V$ is the volume of a set of initial conditions in the 3D state-space of system \eqref{eq:3D-ThomRoss}. The last result shows that volumes in its state-space are preserved in time, and thus it is a volume-preserving dynamical system. We note here that Eq. \eqref{eq:diveregence_condition} is essentially the sum of the diagonal entries in the Jacobian matrix $J$ (see Eq. \eqref{eq:3D-Jacobian}) of system \eqref{eq:3D-ThomRoss}, thus it is equal to its trace. In other words, the divergence of the flow is the trace of the Jacobian matrix of the dynamical system.

As we shall show next, the second property is that system \eqref{eq:3D-ThomRoss} is not force-conservative, in the sense that
\begin{equation}\label{pathclosed}
\oint \overrightarrow{F} .\hskip1mm\overrightarrow{dr}\neq0.
\end{equation}
The \lq\lq{}acceleration\rq\rq{} field $\ddot{X}$, that would be proportional to a force $\mathcal{F}$ (from Newton's second law), can be obtained by computing the time-derivative of Eq. \eqref{eq:3D-ThomRoss}. This yields
\begin{equation*}
\begin{aligned}
\frac{d^{2} X_1}{d t^{2}}&=\sin (X_3) \cos (X_2),\\ 
\frac{d^{2} X_2}{d t^{2}}&=\sin (X_1) \cos (X_3),\\ 
\frac{d^{2} X_3}{d t^{2}}&=\sin (X_2) \cos (X_1),
\end{aligned}
\end{equation*}
which can be written in the compact form
\begin{equation*}
\ddot{X}=\mathcal{F},
\end{equation*}
with $\ddot{X}=(\ddot{X_1},\ddot{X_2},\ddot{X_3})$ and
\begin{equation*}
\mathcal{F}=(\mathcal{F}_1,\mathcal{F}_2,\mathcal{F}_3),
\end{equation*}
where
\begin{equation}\label{eq:mathcal_F1F2F3}
\begin{aligned}
\mathcal{F}_1&=\sin (X_3)\cos (X_2),\\
\mathcal{F}_2&=\sin (X_1) \cos (X_3),\\
\mathcal{F}_3&=\sin (X_2) \cos (X_1).
\end{aligned}
\end{equation}
The potential $U$, if it existed, would be related to the force $\mathcal{F}$ by
\begin{equation*}
\mathcal{F}=-\nabla{U},
\end{equation*}
it would be path independent and would represent the ``mechanical energy'' of the system, when added to its ``kinetic energy''. In this context, $\nabla$ is the gradient of the scalar-valued differentiable function $U: \mathbb{R}^3\rightarrow \mathbb{R}$ that we want to check whether it is the potential energy of system \eqref{eq:3D-ThomRoss}.

\begin{figure}[ht!]
\begin{center}
 \includegraphics[width=0.6\textwidth,height=0.5\textwidth,trim=0 0 0 0, clip]{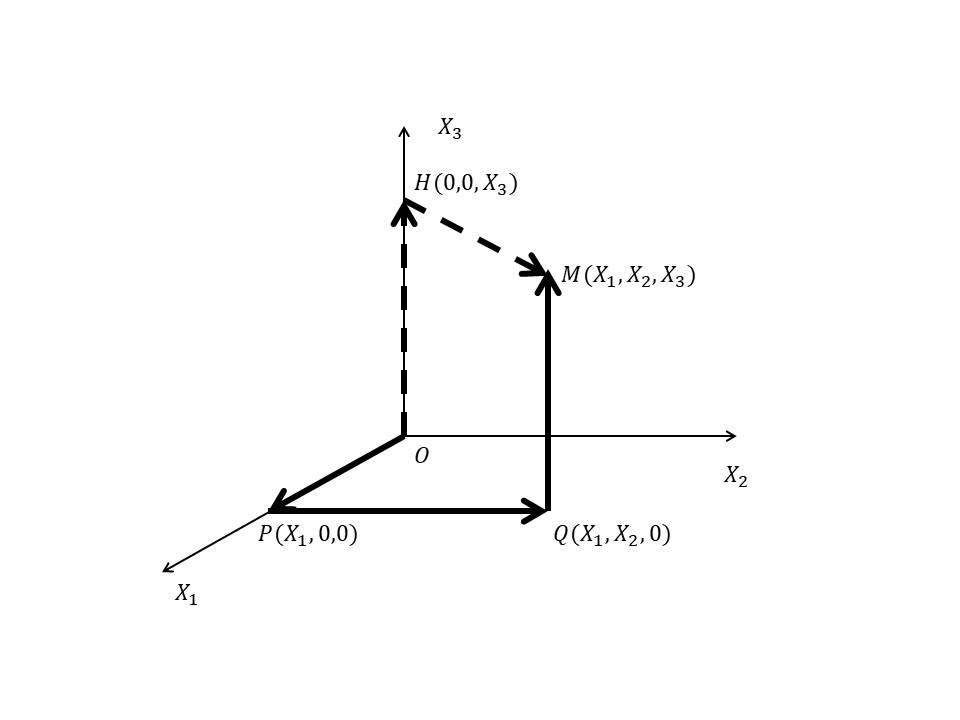}
\caption{An example of two paths (depicted by solid and dashed lines) along which $U$ is different, implying that $\mathcal{F}$ is not a conservative force and thus, the labyrinth walks system \eqref{eq:3D-ThomRoss} is not force-conservative.}\label{two_paths}
\end{center}
\end{figure}

As we show here, this is not the case as we can find a counter-example of two different paths (solid and dashed in Fig. \ref{two_paths}) along which $U$ is not the same function: To this end, let us integrate $\mathcal{F}$ along the two paths shown in Fig. \ref{two_paths}. The solid-line path starts at the origin $O(0,0,0)$ and ends at $M(X_1,X_2,X_3)$, following the three straight lines from $O(0, 0, 0)$ to $P(X_1, 0, 0)$ to $Q(X_1,X_2,0)$ to $M(X_1,X_2,X_3)$ and the dashed-line path, following the straight lines from $O(0,0,0)$ to $H(0,0,X_3)$ to $M(X_1,X_2,X_3)$.

Following the dashed-line path, $U$ reads
\begin{equation*}
\begin{aligned}
U=&-\int_0^{X_1}\mathcal{F}_1(X'_1,X'_2,X'_3)\Bigg\vert_{\substack{X_2\rq{}=0\\ X_3\rq{}=0}}dX_1\rq{}\\&-\int_0^{X_2}\mathcal{F}_2(X'_1,X'_2,X'_3)\Bigg\vert_{\substack{X_1\rq{}=X_1\\ X_3\rq{}=0}}dX_2\rq{}\\&-\int_0^{X_3}\mathcal{F}_3(X'_1,X'_2,X'_3)\Bigg\vert_{\substack{X_1\rq{}=X_1\\ X_2\rq{}=X_2}}dX_3\rq{}.
\end{aligned}
\end{equation*}
Substituting $\mathcal{F}_1$, $\mathcal{F}_2$ and $\mathcal{F}_3$ from Eq. \eqref{eq:mathcal_F1F2F3} into the last equation, $U$ yields
\begin{equation}\label{U1}
U=-X_2\sin(X_1)-X_3\sin(X_2)\cos(X_1).
\end{equation}

Now, following the dashed-line path, $U$ reads
\begin{equation*}
\begin{aligned}
U=&-\int_0^{X_3}\mathcal{F}_1(X'_1,X'_2,X'_3)\Bigg\vert_{\substack{X_1\rq{}=0\\ X_2\rq{}=0}}dX_3\rq{}\\&-\int_{H}^{M}\mathcal{F}\hskip1mm(dX_1\rq{}+dX_2\rq{}),
\end{aligned}
\end{equation*}
which can be written as
\begin{equation*}
\begin{aligned}
U=&-\int_0^{X_3}\mathcal{F}_3(X'_1,X'_2,X'_3)\Bigg\vert_{\substack{X_1\rq{}=0\\ X_2\rq{}=0}}dX_3\rq{}\\&-\int_0^{X_1}\mathcal{F}_1(X'_1,X'_2,X'_3)\Bigg\vert_{\substack{X_3\rq{}=X_3\\ X_2\rq{}=(\frac{X_2}{X_1})X_1\rq{}}}dX_1\rq{}\\&-\int_0^{X_2}\mathcal{F}_2(X'_1,X'_2,X'_3)\Bigg\vert_{\substack{X_3\rq{}=X_3\\ X_1\rq{}=(\frac{X_1}{X_2})X_2\rq{}}}dX_2\rq{}.
\end{aligned}
\end{equation*}
Again, substituting Eq. \eqref{eq:mathcal_F1F2F3} into the last equation, $U$ becomes
\begin{equation}\label{U2}
U=-\frac{{X_1}}{X_2}\sin(X_2)\sin(X_3)+\frac{{X_2}}{X_1}\cos(X_1)\cos(X_3).
\end{equation}

Comparing, $U$ from Eqs. \eqref{U1} and \eqref{U2} computed along the two paths, we deduce they are not the same function. Therefore, $U$ depends on the path, a potential function cannot be defined and system \eqref{eq:3D-ThomRoss} is not force-conservative, since Eq. \eqref{pathclosed} holds. 

This discussion is motivated further by the fact that when one studies the volume-preservation properties of system \eqref{eq:3D-ThomRoss}, given that the trace of its Jacobian matrix is zero (what corresponds to a divergence-free, incompressible flow) and in view of Liouville's theorem for Hamiltonian systems \cite{ODELiouville, AbrahamRalf, Goldstein1980}, one is tempted to consider whether it falls in the realm of Kolmogorov-Arnold-Moser (KAM) theory \cite{KAMArnold}. This might be driven further by the computation of Poincar\'e surfaces of section, such as those in \cite{Arabesque2013,SprottChlouverakis}, where islands of stability can be seen within the chaotic sea and where the diffusive aspect of wandering trajectories is similar in appearance to those of chaotic Hamiltonian systems \cite{Manosetal2012,Skokosetal2004,Skokosetal2007}. 
Yet, this is where similarities stop as labyrinth walks \eqref{eq:3D-ThomRoss}, even though is not a force-conservative system, preserves volumes in its 3D state-space. However, the fact it preserves volumes does not mean there is an integral of motion. The existence of a global conserved quantity in the 3D state-space is prohibited since that would render the system effectively two dimensional and hence chaos could not appear. 
%


\section{Conclusions \& Outlook}\label{Sec:Conclusion}

As all R\"ossler's pioneering contributions, labyrinth chaos still holds promise for very interesting further developments. Its simplicity and elegance, both in terms of symmetries, topology and feedback-circuit structure, makes it a good candidate to compare it with other nonlinear, cyclically coupled systems, such as the Arabesques \cite{ThomRoss2006,Arabesque2013}, the Lotka-Voltera system and its variants \cite{LVgen}, and the Arnold-Beltrami-Childress (1:1:1 ABC) model \cite{ABC111a,ABC111b}.

In connection to arrays of coupled TR systems, where chimera-like states have been shown to emerge even under strong coupling settings, the investigation of the role of inhomogeneities in parameters and in network topology (e.g. scale-free, small-world, Erd\H{o}s-R\'enyi) are worth studying further as well as in connection to other systems \cite{Kaneko2015,Ikedaetal1987}.

Here, we discussed further implications pertaining to the labyrinth walks system by showing that, even though it is volume-preserving, it is not force-conservative. This makes it a fine and elegantly simple example of a low-dimensional volume-preserving, chaotic system with no attractors and with only unstable fixed points. We note that in Statistical Mechanics, where phase spaces are high-dimensional, there are also examples of systems that are volume-preserving, chaotic, with no attractors and with only unstable fixed points, such as extensions of the Nose-Hoover thermostat known as the Martyna-Klein-Tuckerman, chain thermostats \cite{HooverC}.

The investigation of the appropriate Hamiltonian (non-holonomic) or Lagrangian (with constrains) formalism that might encompass the labyrinth walks system, could reveal deeper connections between chaos, volume preservation, symplectic structure, diffusion and KAM theory. It would be interesting to seek a comparison of the labyrinth walks system with the class of systems pertaining to the construction of Hoover thermostats in Statistical Mechanics \cite{HooverA,HooverB,HooverC}. This is because this class of systems provides: (a) a framework for the required analysis by determining phase-space distributions assuming ergodicity and (b) expresses the balance between the kinetic, dissipative and potential parts, what makes their phase-space incompressible.

Finally, information transfer in networks of coupled labyrinth walks systems as ``carriers of information'' \cite{Hiley2002,Hiley1999,vectorpot} emerges as another exciting field of future research, since it is related to the volume-preservation property of individual labyrinth walks systems.

\section{Acknowledgements}
This research was supported by our respective Institutes: Department of Physics, Qom University of Technology for AL, Service de Physique des Syst\`emes Complexes et M\'ecanique Statistique and CeNoLi, Universit\'e Libre de Bruxelles for VB and Department of Mathematical Sciences, University of Essex for CGA.

\section*{Availability of Data}
\noindent The data that support the findings of this study are available within the article.

\bibliographystyle{unsrt}

\end{document}